\documentclass[aps,prb,groupedaddress,twocolumn,10pt,showpacs]{revtex4-1}
\usepackage[utf8]{inputenc}
\usepackage{graphicx}
\usepackage{amsmath,amssymb,color,subfigure,braket,bbm}
\usepackage{pgf}
\usepackage{tikz}
\usepackage{mathcomp}
\usepackage[capposition=bottom]{floatrow}
\usepackage{natbib}

\begin{document}

\title{Fermi-surface Reconstruction in the Repulsive Fermi-Hubbard Model}
\author{Ian Osborne$^{1}$}
\author{Thereza Paiva$^{2}$}
\author{Nandini Trivedi$^{1}$}

\affiliation{(1) Department of Physics, The Ohio State University, Columbus, OH  43210, USA}
\affiliation{(2) Instituto de F\'\i sica, Universidade Federal do Rio de Janeiro, Caixa Postal 68.528, 21941- we 972, Rio de Janeiro, RJ, Brazil}

\date{\today}


\begin{abstract}
One of the fundamental questions about the high temperature cuprate superconductors is the size of the Fermi surface (FS) underlying the superconducting state. By analyzing the single particle spectral function for the Fermi Hubbard model as a function of repulsion $U$ and chemical potential $\mu$, we find that the Fermi surface in the normal state reconstructs from a large Fermi surface matching the Luttinger volume as expected in a Fermi liquid, to a Fermi surface that encloses fewer electrons that we dub the ``Luttinger Breaking" (LB) phase, as the Mott insulator is approached. This transition into a non-Fermi liquid phase that violates the Luttinger count, is a continuous phase transition at a critical density in the absence of any other broken symmetry. We obtain the Fermi surface contour from the 
spectral weight $A_{\vec{k}}(\omega=0)$ and from an analysis of the poles and zeros of the retarded Green's function $G_{\vec{k}}^{ret}(E=0)$, calculated using determinantal quantum Monte Carlo and analytic continuation methods.
We discuss our numerical results in connection with experiments on Hall measurements, scanning tunneling spectroscopy and angle resolved photoemission spectroscopy.
\end{abstract}

\maketitle

\noindent {\it Introduction:}
A question of fundamental importance for strongly correlated metals near a Mott transition is: What is the size of the Fermi surface (FS)? Does it count all the electrons or only the carriers relative to the Mott filling? In other words, is the Fermi surface large or small\cite{Luttinger}?
And furthermore, if the FS deviates from the Luttinger volume and there is Fermi surface reconstruction, is it due to competing order or due to topological order? 


We are motivated by three sets of experiments on the cuprates: the Hall coefficient which gives information on the density and type of carriers, scanning tunneling spectroscopy that gives information about broken charge density and pair density order, and angle resolved photoemission spectroscopy that gives information about the momentum resolved density of states. The Hall number $n_H$ in YBa$_2$Cu$_3$O$_y$ (YBCO) shows a distinct change at a critical doping $p_c^H$ from $n_H\approx 1+p$ at high doping $p$ of holes to $ n_H \approx p$ for low doping. \cite{ChangeOfCarrierDensityAtThePseudogapCriticalPoint-Tallifer} The next question is whether the change in behavior of the Hall coefficient occurs due do broken symmetry in the charge, spin or pairing channel. 
Scanning tunneling spectroscopy experiments indicate that charge order is observed in YBCO below a critical doping $p_c^{CDW}< p_c^H$, which suggests that the mechanism causing the change of the Hall coefficient and the mechanism driving charge order are distinct phenomena.

Here we sharpen the question for the celebrated Hubbard model rather than focusing on analysis of experiments; the latter being undoubtedly more complicated.
Early QMC calculations of the spectral function related antiferromagnetic fluctuations to pseudogap formation and quasiparticle weight transfer~\cite{Bulut, Preuss, Moreo}. More recent work focused on systems with particle-hole asymmetry, introduced by next-near neighbour hopping. Cluster Dynamical Mean-Field (CDMFT) studies have shown that the quasi-particles show momentum-dependent renormalizations due to proximity to the Mott transition, even in the absence of long-ranged antiferromagnetic correlations.~\cite{Civelli,Gull}

In this paper, our aim is to extract the underlying FS as a function of doping, with particular emphasis on the region close to the Mott transition.\cite{Shastry}
We focus on the particle-hole symmetric Hubbard model with only nearest-neighbor hopping. Our main result is that the FS volume follows the Luttinger volume for high densities, but starts deviating below a critical density $n_c$ as the Mott density is approached, as shown in Fig.~1. While such ideas have been discussed in the literature previously, what is significant is that we present the first quantum Monte Carlo results on Fermi surface reconstruction with no systematic errors. We show that FS reconstruction does not occur abruptly from a volume that counts $(1+p)$ holes to $p$ holes, but shows a continuous evolution below $n_c$. A similar evolution occurs on the hole- doped side as well.

We calculate the FS contour by using determinantal quantum Monte Carlo methods to obtain the imaginary time Green function, $G_{\vec{k}}(\tau)$, as a function of the interaction strength, carrier concentration, and temperature. Analytic continuation of $G$ yields the spectral function $A_{\vec{k}}(\omega)$ \cite {MaxEnt-Gubernatis, MaxEnt-Sandvik, MaxEnt-Jarrell, MaxEnt-Bouadim} whose contour at $\omega=0$ then yields the FS. 

We also analyze the behavior of the retarded Green's function, $G_{\vec{k}}^{ret}(E)$ and find that it changes sign at zero chemical potential in two distinct ways: (i) For densities away from the Mott density of one particle per site, the sign change occurs through a pole, as expected for a system with well-defined quasiparticles; (ii) As the Mott density is approached, the sign change occurs through a zero. Such a behavior was first pointed out by Dzyaloshinskii~\cite{D-zeros} as occurring in a Mott insulator. In this paper, we find remarkably that the breakdown of the Luttinger count happens in the metallic state approaching the Mott insulator and the retarded Green's function, $G_{\vec{k}}^{ret}(E)$ changes sign at $\mu=0$ through a zero in the function. Notably, this Fermi surface reconstruction is found to occur in the absence of any competing order, indicating the emergence of a ``Luttinger Breaking" (LB) non-Fermi Liquid phase, a signature of topological order close to the Mott transition~\cite{D-zeros,oshikawa}.

\begin{figure}
\includegraphics[width = 8.6 cm]{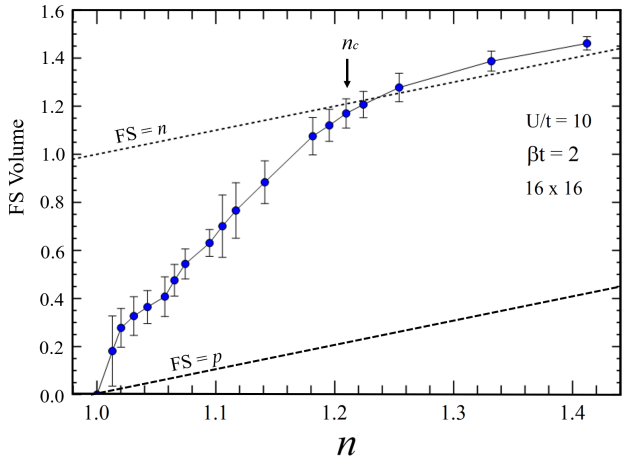}
\caption{
Evidence for Fermi Surface Reconstruction: The Fermi surface volume deviates strongly from the expected Luttinger volume which is proportional to the charge carrier density, below a critical density $n_c$ as the Mott insulator at $n=1$ is approached. Determinantal QMC results for the Hubbard model on a 16 by 16 square lattice for $U/t = 10t$ and $\beta t = 2$ showing $n_c \approx 1.2$.
}
\label{fig:compressibility}
\end{figure}

\medskip

\noindent {\it Models and Methods:}
The Fermi Hubbard model is the paradigmatic model for Mott insulators, strongly correlated metals, and high temperature superconductors. 
In its particle-hole symmetric form, the Hamiltonian is given by

\begin{align}\label{eq:Hamiltonian}
\begin{aligned}
    \mathcal{H} = -t& \sum_{\langle i,j \rangle,\sigma}{ \left( \hat{c}^\dag_{i,\sigma} \hat{c}_{j,\sigma} + h.c. \right) }\\
    &+ 
    U ~\sum_{i}{ \left( \hat{n}_{i\uparrow} - \frac{1}{2} \right) \left( \hat{n}_{i\downarrow} - \frac{1}{2} \right)}
    - \mu \hat{N}
\end{aligned}
\end{align}
defined so that when the chemical potential $\mu = 0$, the average density is unity ensuring that the system is half-filled. Here $t$ is the tunneling amplitude for a fermion to hop from one site to a neighbor without changing the spin, $\sigma= \uparrow, \downarrow$, and $U$ is the on-site Coulomb repulsion.  
The spatial index $i$ labels a site on a 2D square lattice, and 
 $\hat{c}_{i,\sigma}$ and $\hat{c}^\dag_{i,\sigma}$ are fermionic annihilation and creation operators respectively.
The number operator is defined as $\hat{n}_{i,\sigma} \equiv \hat{c}^\dag_{i,\sigma} \hat{c}_{i,\sigma}$,  $\hat{n}_i = \hat{n}_{i, \uparrow} + \hat{n}_{i, \downarrow}$, and the particle density per site ${n} = \sum_i{\langle\hat{n}_i\rangle}/{N_s}$, where $N_s$ is the total number of sites.
Relative to the filled band with two electrons per site, the hole density is $1 + p$. 
Particle-hole symmetry is exhibited by the transformation of particle creation and annihilation operators for hole annihilation and creation operators respectively: $c^\dag_{\sigma i} \xrightarrow{} (-1)^i d_{\sigma i};~ c_{\sigma i} \xrightarrow{} (-1)^i d^\dag_{\sigma i}$.
A finite next nearest neighbor hopping term, $t'$, which we do not include in our discussion here, would break particle hole symmetry in Eq. \ref{eq:Hamiltonian}.

We next calculate thermodynamic properties and the single-particle Green function by implementing the determinantal Quantum Monte Carlo (QMC) algorithm which essentially employs a Trotter-Suzuki decomposition to break up the non-commuting hopping and interaction terms in imaginary time $\tau$.
\cite{Blankenbecler,Raimundo}
This maps the original 2D quantum problem to a (2+1)D classical problem where the extra dimension is set by the inverse temperature $\beta=1/{(k_B T)}$.
Details of the calculations and a discussion of the sign problem \cite{sign-1, sign-2, Raimundo} can be found in the supplementary material.

\begin{figure*}[t]
    \includegraphics[width = 17.2cm]{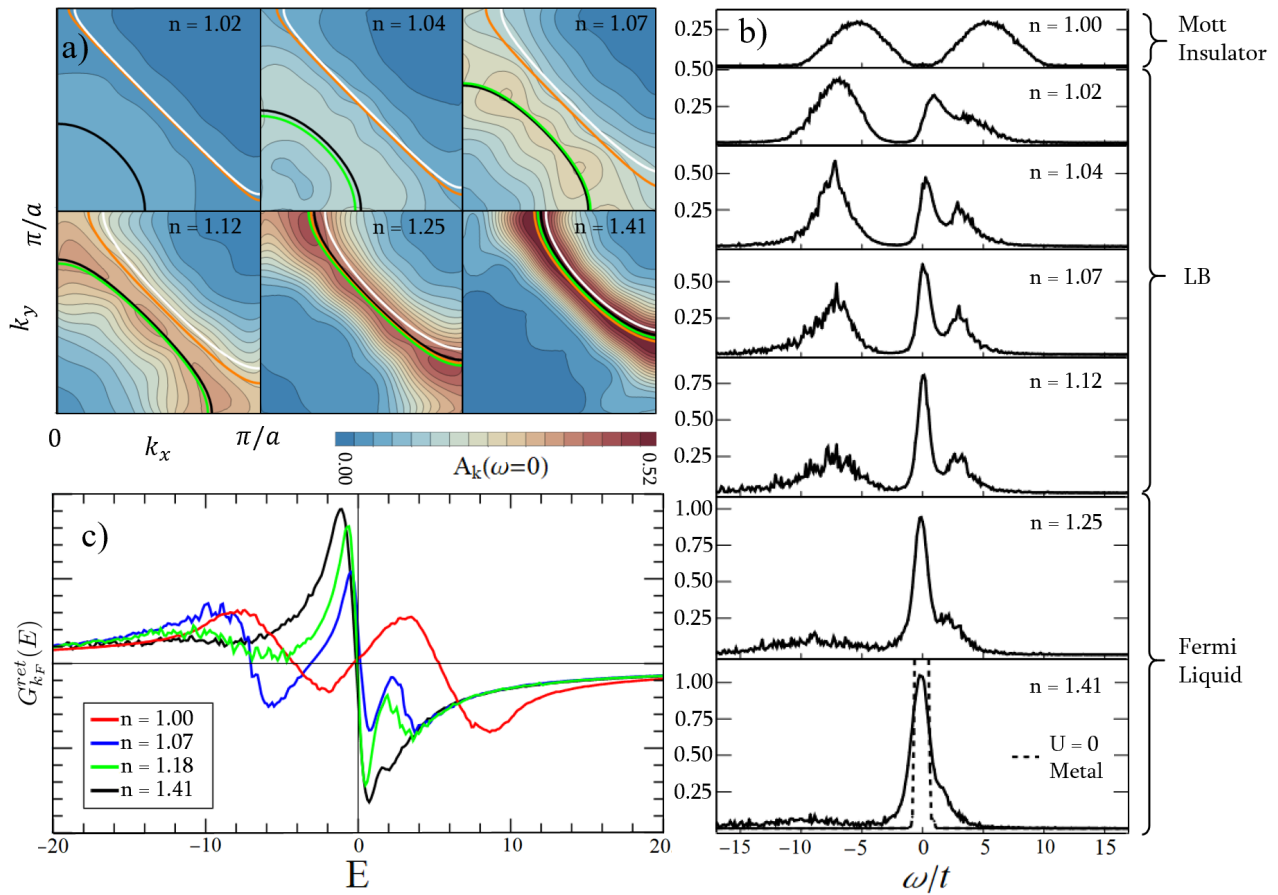}
    \caption{Spectral Functions of Hubbard model:
    a)  Fermi surface contours for a square lattice Brillouin zone calculated by 4 different methods for the Hubbard model at $U=10t$ as a function of density. (i) Luttinger's Theorem applied to a free system (white); Momentum distribution function (MDF) contour formed by $ n_{\vec{k}}=1/2$(orange); (iii) Spectral weight at the Fermi energy $A_{\vec{k}}(\omega=0)$ (black); and (iv) singularity of retarded Green's function $G_{\vec{k}}^{ret}(E=0)$ (green). Lattice size 16 $\times$ 16 site square and $\beta t = 2$. 
    b) Spectral function $A_{\vec{k}}(\omega)$ averaged over k-states on the Fermi surface contour. 
    The total spectral weight is normalized $\frac{1}{\pi}\int_{-\infty}^{\infty}{A_{\vec{k}}(\omega) d\omega} = 1$. For comparison the non-interacting metal spectral function is shown at the same temperature, and on the same lattice size (dashed) and contrasted with the interacting metal for $n=1.41$. 
    ``Fermi Liquid" is distinguished from the ``Luttinger Breaking" (LB) regime by the agreement of the MDF and spectral contours. The Mott insulator occurs at n=1.
    c) The retarded Greens function is calculated using equation \ref{eq:Retarded Greens} for the spectral functions shown in \ref{Spectral Function}b. A pole-like sign change of the Green's function at $E=0$ indicates the presence of a quasiparticle on the Fermi surface, consistent with the behavior in a Fermi liquid. The behavior changes dramatically in the Luttinger Breaking regime with sign changes at finite energy and a zero of the Green function at zero energy.}
    \label{Spectral Function}
\end{figure*}

\medskip
\noindent {\it Restructuring of the Fermi surface:}
From QMC we directly calculate the Green function in imaginary time $\tau$ and from that using an analytic continuation procedure  we obtain the spectral function:

\begin{equation} \label{eq:Spectral function}
    G_{ \vec{k}}(\tau) = \int_{-\infty}^{\infty}d\omega \left [  \frac{e^{-\omega \tau}}{1 + e^{-\beta \omega}}\right ] A_{ \vec{k}}(\omega) 
\end{equation}

The spectral function $A_{ \vec{k}}(\omega)=-(1/\pi) {\rm Im} G^{ret}_{ \vec{k}}(\omega)$ gives information about the probability of finding an electron in state $(\vec{k},\omega) $.  
Fig. \ref{Spectral Function}(a) shows the contour of the spectral function $A_{\vec{k}}(\omega=0)$ over the Brillouin zone, and the locus of the spectral weight (black curve).\cite{AGDBook}
A related quantity, the retarded Green's function, defined by, \begin{align} \label{eq:Retarded Greens}
    G_{\vec{k}}^{ret}(E) = {\cal P}\int_{-\infty}^{\infty}{d\omega \frac{A_{\vec{k}}(\omega)}{\omega -E}},
\end{align}
provides a second indicator of the Fermi surface contour as the singularities of $G_{\vec{k}}^{ret}(E=0)$ mark the boundary where the sign change occurs (shown in green)~\cite{D-zeros,philip-philips}. Here ${\cal P}$ denotes the principal part of the integral. Note that while $A_{\vec{k}}(\omega=0)$ uses only information at $\omega=0$ to map the Fermi surface contour, $G_{\vec{k}}^{ret}(E=0)$ uses information over the entire spectral range of $A_{\vec{k}}(\omega)$ to map the contour. 

\begin{figure*}
\includegraphics[width =16cm]{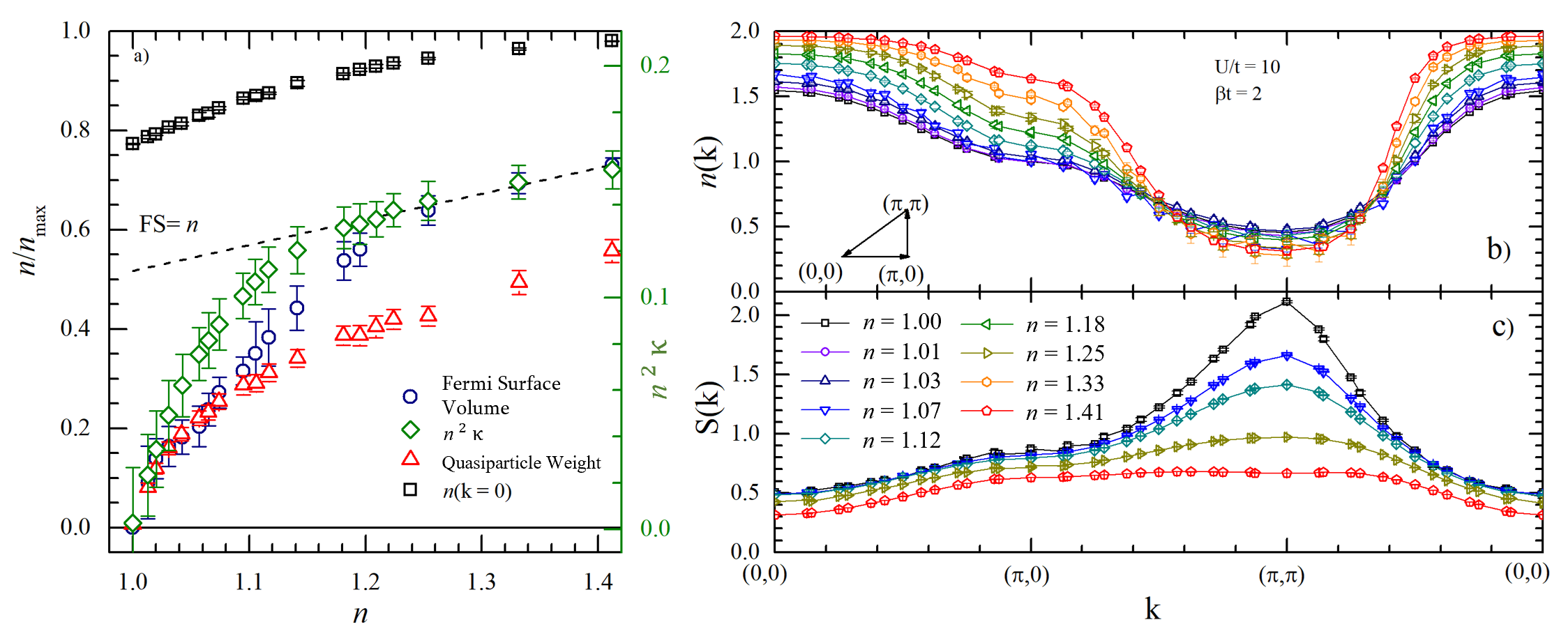}
\caption{
a) Behavior of (i) the occupation of the $n_{\vec{k}}=(0,0)$ state, (ii) the quasiparticle weight around the chemical potential, (iii) the compressibility $n^2 \kappa=dn/d\mu$, and, (iv) The FS volume (also shown in Fig.~\ref{fig:compressibility}), all plotted as a function of density.
b) Momentum distribution function $n(\bf k)$ as a function of density for multiple values of $n$. 
c)  Structure factor across the Brillouin zone for multiple values of $n$ obtained from the spin-spin correlations showing the absence of long range magnetic ordering. 
Lattice size $16\times 16$, $U=10t$, $\beta t=2$. 
}
\label{fig:orderParameter}
\end{figure*}

Luttinger's theorem asserts that the volume enclosed by the Fermi surface of an interacting Fermi liquid is proportional to the number of particles in the system. \cite{Luttinger}
This allows us to find the reference non-interacting Fermi surface corresponding to the actual density obtained by QMC for the specific set of parameters $(U,T,\mu)$ (white contour in Fig.~\ref{Spectral Function}(a)).
It is also useful to compare with the contour obtained from the momentum distribution function (MDF) (Fig. \ref{fig:orderParameter}(b))
$ n_{\vec{k}}=1/2$ calculated by QMC (shown in orange in Fig. \ref{Spectral Function}(a)). In the thermodynamic limit for a non-interacting system at $T=0$, the MDF has a jump of size unity $Z=1$ at the Fermi wave vector $k_F({\vec k})$ when the system transitions from occupied states below $k_F$ to zero occupancy above. In a Fermi liquid, following Luttinger's theorem, $k_F({\vec k})$ does not change and $0<Z\le 1$. Due to inter-electron interactions, some of the states below $k_F({\vec k})$ are scattered into states above but nevertheless in a Fermi liquid, a finite step at $k_F$ persists at $T=0$. At finite $T$, naturally the step gets rounded; however, from the peak in the gradient of the MDF (Fig. \ref{fig:orderParameter}(b)) as a function of $\vec k$, we can extract the location of the underlying FS.

For dopings greater than $p\gtrsim 0.2$ each of these methods of finding the FS show stark agreement.
Such a validity of Luttinger volume is found even for large $U$ for sufficiently large doping.
However, when the doping is less than $p\lesssim 0.2$, we observe a departure of the FS contours obtained from these four methods.
The size of the orange and white surfaces, corresponding to the $n_{\vec{k}} = 1/2 $ contour and the Luttinger surface respectively, count the total electronic density. 
The spectral weight and retarded Green's function boundaries, black and green contours respectively, on the other hand, recede to include fewer states.
In other words, the spectral function methods indicate that the Fermi surface reconstructs from a large Fermi surface enclosing $n = 1 + p$ fermions to a smaller one as quantified in Fig. \ref{fig:compressibility}, below a critical density of $n_c\approx 1.2$. 
As observed in Fig. \ref{Spectral Function} (a), the black and green contours transform from being close to a diamond shaped Fermi surface to a small circular Fermi surface centered around the $\Gamma$ point; the deviation occurs for densities $n\lesssim 1.2$. 

The quasiparticle weight, shown in Fig \ref{fig:orderParameter} (a), shows the fraction of the spectral function, $A_{\vec{k}_f}(\omega),$ around zero energy, defined by $ 
    QW = \frac{1}{\pi}\int_{-\epsilon}^{\epsilon} {A_{\vec{k}_f} (\omega) d\omega}$
where as a reference we use the $U=0$ spectral function of the finite size broadened non-interacting metal from a delta function to a Lorentzian distribution of width $2 \epsilon$ in Fig \ref{Spectral Function} (b) (lowest panel) to account for the resolution in the analytic continuation procedure.
The interacting system shows the development of incoherent side bands or Mott bands around the peaked spectral function at $\omega=0$. The Fermi surface restructuring is already visible below $n\approx 1.2$ and the deviation of the actual Fermi contour from the Luttinger contour only gets more pronounced as the incoherent weight increases upon approaching the Mott transition at $n=1$.

It is important to note that there is no evidence of long range anti-ferromagnetic order at the temperatures and parameters we are analyzing the Fermi surface. The spin structure factor at $(\pi, \pi)$, shown in Fig \ref{fig:orderParameter} (c), shows a small peak at $n=1$ which gets quickly suppressed as the density moves away from this commensurate value. 

\medskip

\noindent {\it Discussion and Outlook:}

Our results suggest that if secondary ordering in the spin, charge, or pairing channel is suppressed, the proximity to a Mott insulator alone drives the reconstruction of the Fermi surface to a ``Luttinger Breaking" (LB) phase. Featureless doped quantum spin liquids are perhaps the most promising platform for the observation of such a state of matter. Sachdev, Senthil and collaborators have proposed the possibility of a non-Fermi liquid phase (dubbed FL$^\ast$) with a Fermi surface composed of fractionalized spinons with a volume $(n - 1)\mod 2$. \cite{FractionalizedFermiLiquids}
They claim that quantum fluctuations of the antiferromagnetic order parameter generate emergent gauge fields that lead to a new state of matter with topological order~\cite{SachdevPNAS,SachdevPRX,SachdevarXiv}. Further diagnostics on the entanglement properties of these non-Fermi Liquid phases are required to understand the connection between our discovery of the LB and the proposed FL* phases.

A topological framework~\cite{D-zeros,seki-yunoki} for understanding the transition from a FL obeying the Luttinger count to the LB phase that violates this count, is obtained by expressing the Luttinger volume as the winding number of the single-particle Green’s function at finite
temperatures. Further, the winding number can be connected with the distribution of quasiparticles and the Luttinger volume and it can be shown non-perturbatively that for a strongly interacting Hamiltonian that preserves particle-hole symmetry both types of behavior, pole and zero, of the Green function at zero energy are observed.

We do find that the LB phase with a reconstructed Fermi surface provides a natural description of the pseudogap phase. Our results on the Hubbard model are the first indication of such an LB phase from a controlled calculation with only statistical errors. Going forward, further simulations are necessary to study an extended Hubbard model with nearest neighbor hopping to describe the more realistic parameters for the cuprates. Here our aim was to show the emergence of the LB phase in the very simplest case with only nearest neighbor hopping. It would also be interesting to push the calculations to even lower temperatures to see the emergence of the superconducting phase from the LB phase for doping values $p$ below $p_c^{PG}$ and contrast that with the superconducting phase that emerges from the FL phase above this critical value.

\medskip

\noindent {\it Acknowledgments:} We acknowledge useful discussions with Hasan Khan during the initial stages of this project. We also acknowledge insightful discussions with Mohit Randeria and Sumilan Banerjee. I.O. and N.T. acknowledge the support of the DOE-BES Grant No. DE-FG02-07ER46423. T. P. acknowledges support from CNPq, FAPERJ and INCT on Quantum Information.

\bibliographystyle{apsrev4-1}
\bibliography{references}

\end{document}


\title{Supplement to Topological Fermi-surface Reconstruction in the Repulsive Fermi-Hubbard Model}
\author{Ian Osborne$^{1}$}
\author{Thereza Paiva$^{2}$}
\author{Nandini Trivedi$^{1}$}

\affiliation{(1) Department of Physics, The Ohio State University, Columbus, OH  43210, USA}
\affiliation{(2) Instituto de F\'\i sica, Universidade Federal do Rio de Janeiro, Caixa Postal 68.528, 21941-972, Rio de Janeiro, RJ, Brazil}

\date{\today}

\maketitle

\section{Details about the simulations}

\begin{figure}[t]
\includegraphics[width =8.6cm]{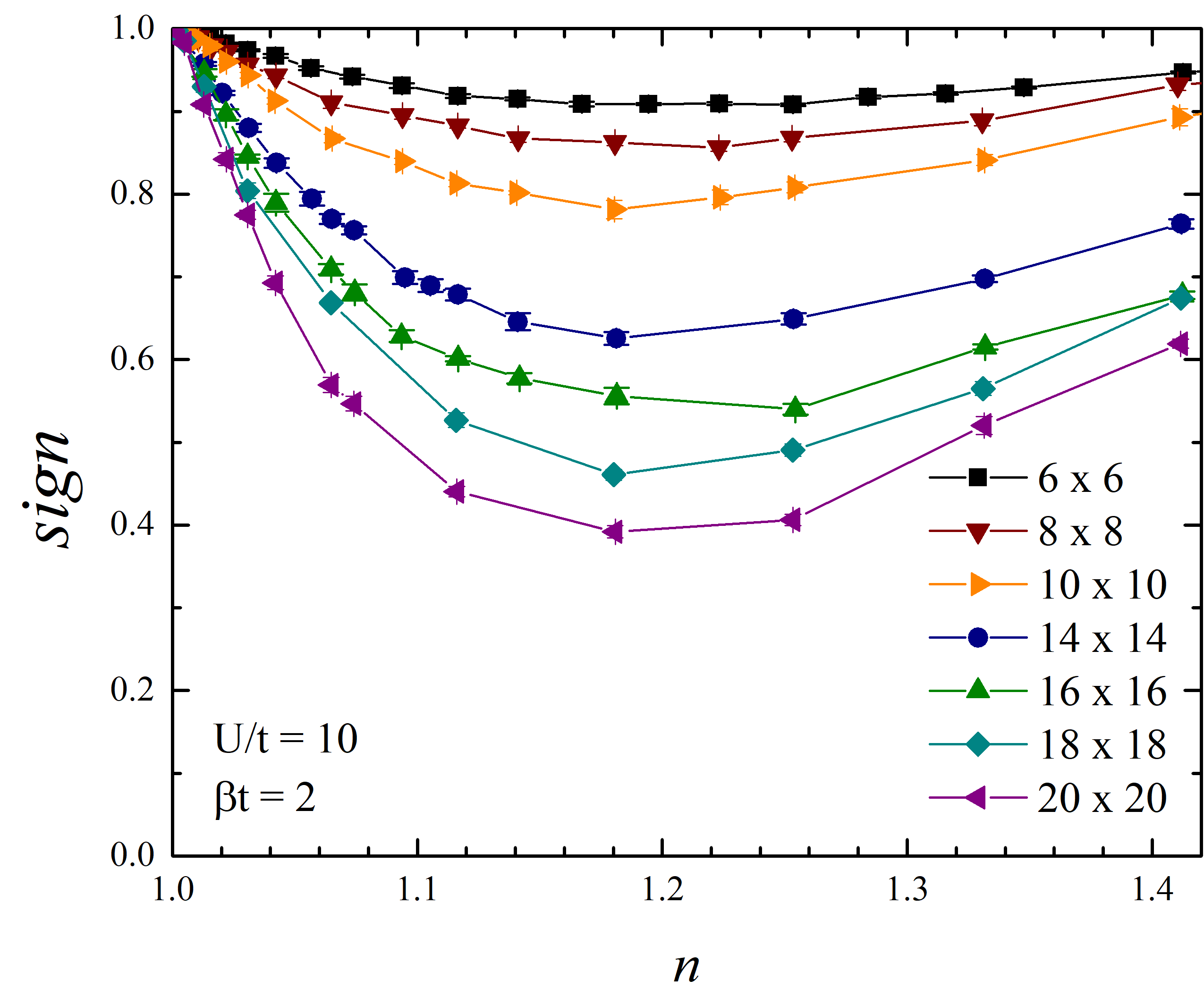}
\caption{
a) Average total sign as a function of density for lattice sizes from $6 \times 6$ to $20 \times 20$.
Temperature is set such that $\beta t= 2$, and the interaction potential is $U = 10t$. 
}
\label{fig:sign}
\end{figure}

In Determinantal Quantum Monte Carlo (DQMC) the grand partition function is expressed as a sum over all Ising spin configurations $c \equiv \{s\}$ at each space-time lattice point, of a product of determinants. 
\begin{align}
{\cal Z}= \Big( \frac{1}{2}\Big)^{L^d M} {\mathbf {\mathrm {\Large Tr}}}_ {\{s\}} \  {\rm det} O^\uparrow(\{ s\} ) \cdot {\rm det} O^\downarrow(\{s\}) 
\end{align}
where $M=\beta / \Delta \tau$, $L$ is the linear size of the system and  $d$ the dimension. The ``Boltzmann weight" is given by the product $p(c)={\rm det} O^\uparrow(\{ s\} ) \cdot {\rm det} O^\downarrow(\{s\})  $ which is not always positive. We can keep track of the sign by writting
\begin{align}
p(c)= sign(c) \  |p(c)|,   
\end{align}
where $sign(c)= \pm 1$, this way the absolute value is used as the weight in the Monte Carlo procedure and the sign is included in the measurements. 
Any expectation value $\langle A \rangle$ is then given by
\begin{align}
\langle A \rangle = \frac{ \sum_c p(c)A(c)}{\sum_c p(c)}=\frac{ \sum_c |p(c)| sign(c) A(c)}{\sum_c |p(c)| sign(c)} \equiv \frac {\langle sign \ A \rangle }{\langle sign \rangle}.   
\end{align}
At low temperatures, both  $\langle A  \ sign \rangle$ and  $\langle sign \rangle$ become very small, leading to the well known ``fermion sign problem". \cite{sign-1, sign-2, Raimundo}
The fermion sign problem is also known to get worse with increasing system size as is shown in figure \ref{fig:sign}. 
We have then restricted our system sizes to lattices up to $16 \times 16$, where $\langle sign \rangle > 0.5$.

\begin{figure}[b]
\includegraphics[width = 8.6cm]{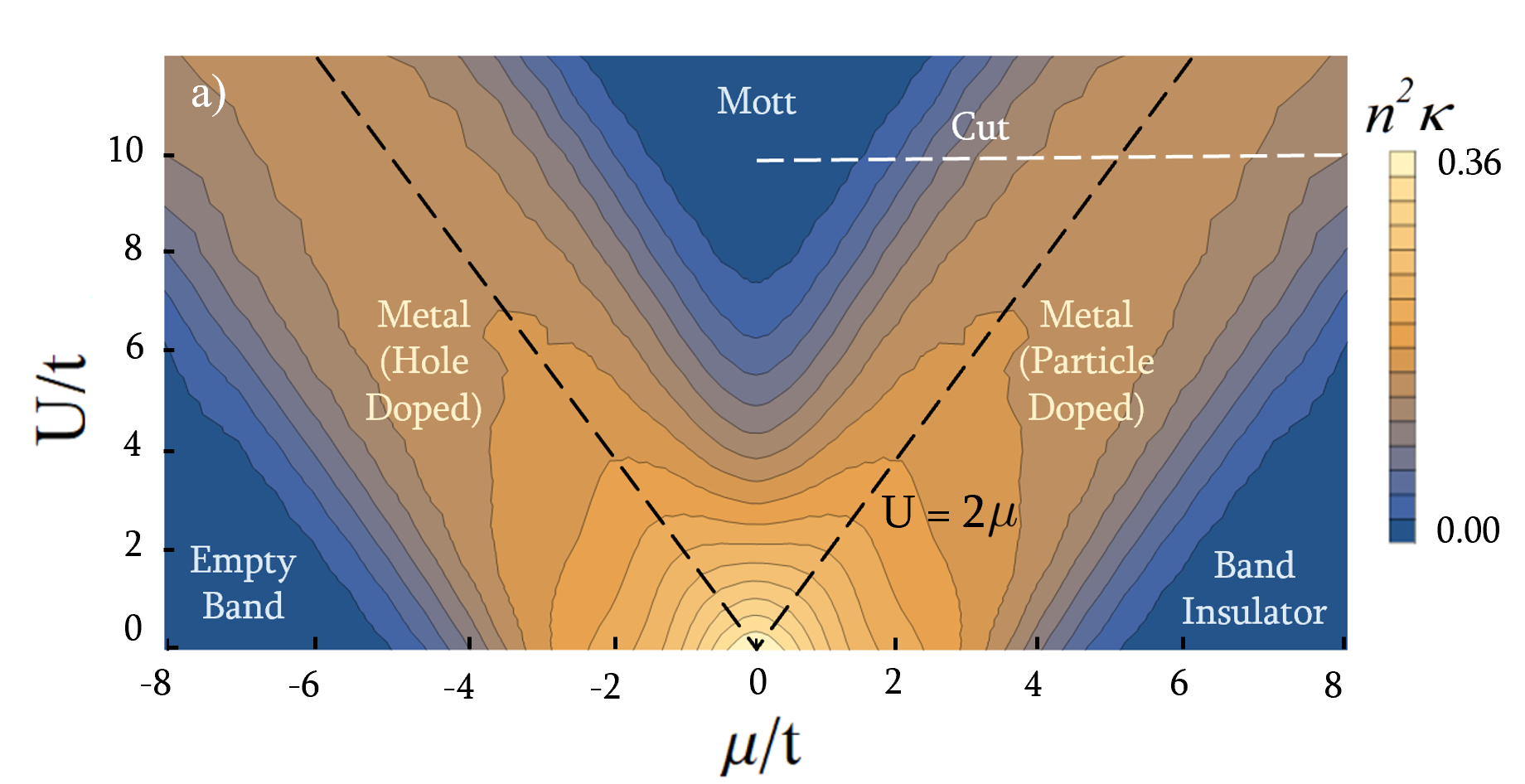}
\caption{
Compressibility $\kappa n^2$ is calculated on a 6 $\times$  6 lattice as a function of interaction potential (U) and chemical potential ($\mu$) at $\beta t = 2$. Small (close to zero) compressibility (blue) indicates Mott and band insulating phases; positive compressibility corresponds to a metallic phase. 
}
\label{fig:phaseDiagram}
\end{figure}

One sweep of DQMC constitutes the metropolis algorithm attempting to flip the Ising spin on every site on the square lattice and for every time slice.
1000 warm up sweeps are sufficient to reach equilibrium, while 15000 sweeps  are used for measuring thermodynamic quantities with the associated error bars.
For the (2 + 1)D lattice, we use $\Delta \tau =1/40$ and  $M=80$ imaginary time slices, which gives $\beta t =M \Delta \tau=2$ and a temperature $ T = 0.5 t$ .
The interaction potential is held constant at an order of magnitude greater than $t$: $U = 10 t$.

We average the results from seven trials trials to increase statistics.
Four of these trials are conducted on a 16 $\times$ 16 cluster,
three on a 14 $\times$ 14 cluster. 
The two lattice sizes compliment by providing data for k-states that are in-accessible by either lattice size individually. 
The different lattice sizes affect quantities like density and compressibility to a negligible degree.

\section{Compressibility}

\begin{figure}[t]
\includegraphics[width = 8.6cm]{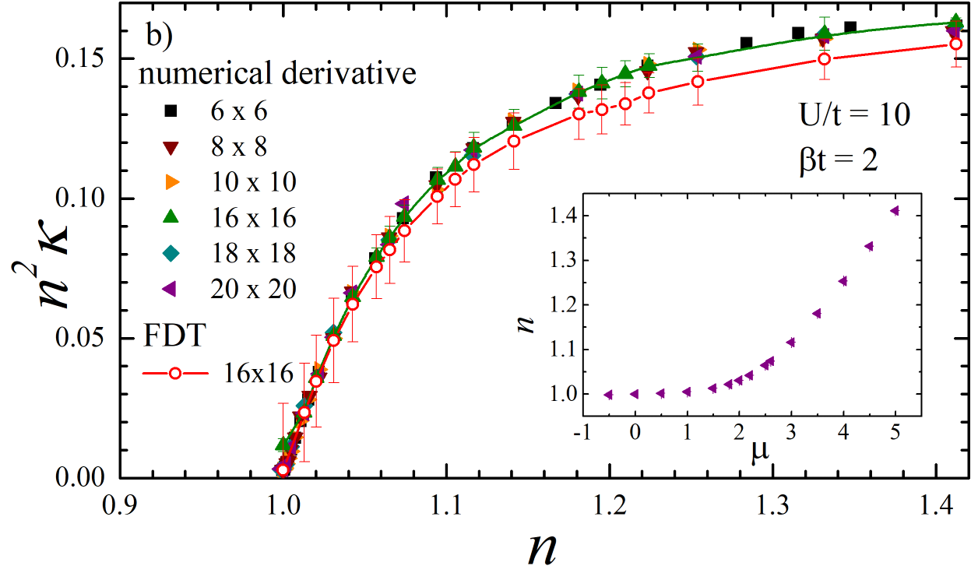}
\caption{
Compressibility $\kappa n^2$ is calculated for lattice sizes from 6 $\times$ 6 to  16 $\times$ 16 system with a fixed interaction strength $(U$ = 10$t$) and varying density/ chemical potential ($\mu$) for $\beta t = 2$. Closed symbols are data extracted from numerical derivatives $dn/d\mu$ and open symbols from fluctuation dissipation theorem.
}
\label{fig:FDT}
\end{figure}

The compressibility $\kappa(U,\mu,T)$ is a useful diagnostic of the phases as it is zero in the Mott and band insulating phases (at $T=0$) but non-zero in metallic and superconducting phases as shown in Fig \ref{fig:phaseDiagram}.
$\kappa n^2$ is obtained from the fluctuation dissipation theorem:
\begin{equation}
        \kappa  n^2  =\frac{1}{N_s} \frac{d \langle \hat{N} \rangle }{d \mu}=\frac{\beta}{N_s} \left( \langle \hat{N}^2 \rangle - \langle \hat{N} \rangle ^2 \right)
\end{equation}
where $\langle \hat{N}^2 \rangle$ is given in terms of correlation functions $ \left\langle \sum_{i,j}{\hat{n}_i \hat{n}_j} \right\rangle $
calculated directly using QMC techniques. 

At half-filling, strong Coulomb interaction ($U/t\gg 1$) drives a phase transition from a metallic state to a Mott insulator (Fig \ref{fig:phaseDiagram}) at a finite temperature $T$.  At a fixed interaction strength, a Mott insulator to metal transition is driven by tuning the chemical potential, as seen by the behavior of $\kappa$ in Fig \ref{fig:FDT}; the agreement between $\kappa$ obtained by fluctuation- dissipation  and the numerical derivative of the density with respect to $\mu$ is a good check of our numerics. 

For a finite $U$, at $n=1$ the system is incompressible with $\kappa=0$, as expected in a Mott insulator. Away from half filling in the metallic phase, the effect of strong interactions continues to dominate and is evident in the strong suppression of $\kappa$ as shown in Fig. 
\ref{fig:FDT}.
The low compressibility can be understood as arising from a background of ``Mott" sites (1 particle on every site) that essentially remain frozen and do not contribute to $\kappa$. Only the excess doubly occupied sites (doublons) contribute to $\kappa$. 

\begin{figure}[b]
\includegraphics[width =8.6cm]{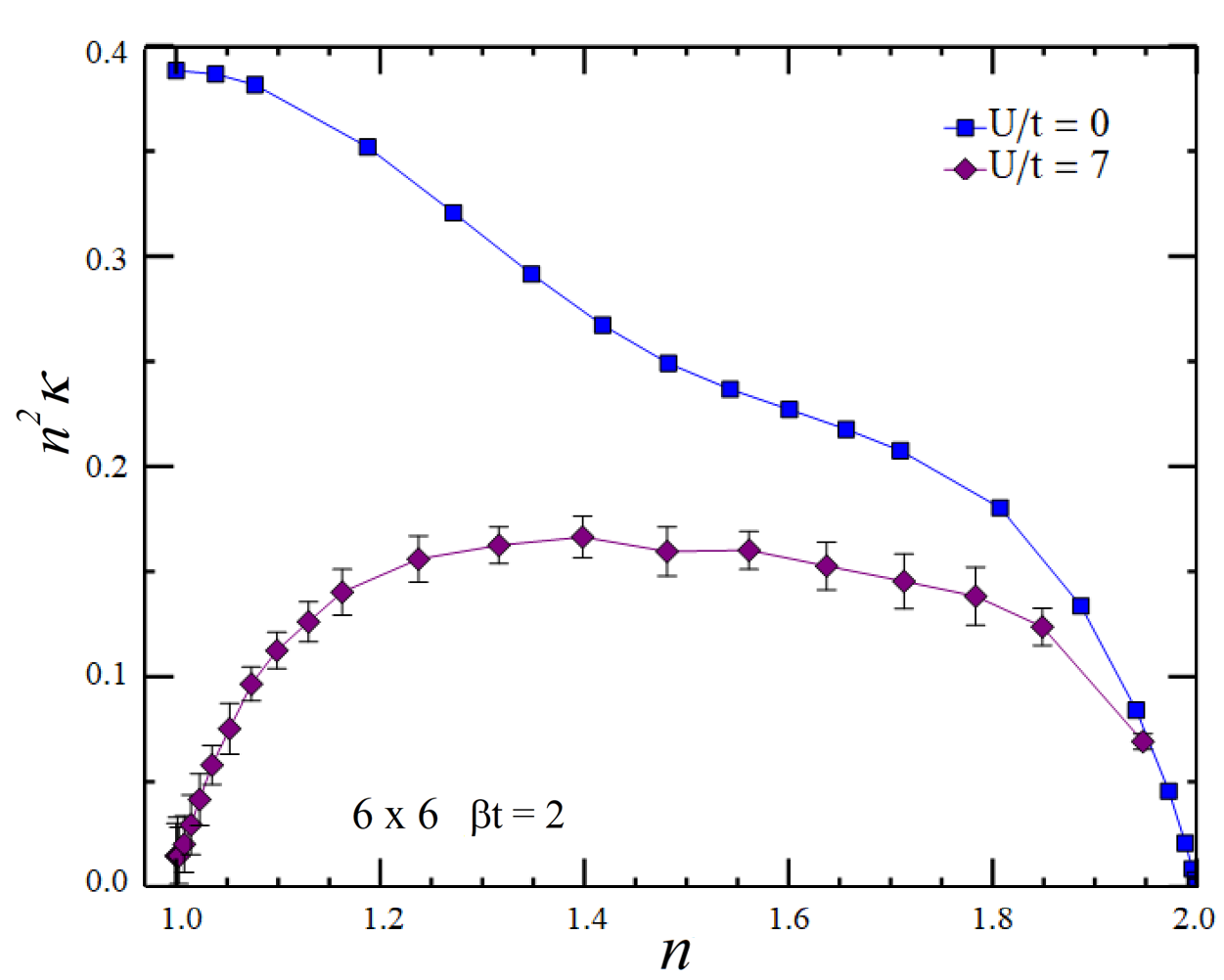}
\caption{
Compressibility for a strongly interacting system and a non-interacting system, at $T=t/2$ and for a $6 \times 6$ lattice.
}
\label{fig:Noninteracting compressibility}
\end{figure}

In Fig. \ref{fig:Noninteracting compressibility} the compressibility of a non-interacting ($U/t = 0$) and interacting ($U/t = 7$) system are compared for a cluster size of 6 $\times$ 6 and temperature of $T= t/2$.

We posit that at large interaction values there is a background of Mott sites, on which the charge carriers are essentially the doublons or holes depending on whether the density is higher than unity or lower, with a density of $p   = |n-1|$. 
Using this assumption, we notice that for large interaction strengths, around half filling the compressibility is suppressed due to a lack of charge carrying doublons or holes and the system becomes a Mott insulator.
For fillings close to 2, the compressibility should approach zero as the band gets fully filled and the system becomes a band insulator.

Both of these claims are confirmed by Fig. \ref{fig:Noninteracting compressibility}.
Another non-trivial check of our assumption is that the compressibility of the strongly-interacting system at $n = 1+p = 1.5$ should be approximately half the compressibility of the non-interacting system at $n = 1$: $\kappa(U/t\gg 1, p = 0.5) \approx \frac{1}{2} \kappa(U/t = 0, p = 0) $.
This is because the approximate density of charge carriers in the non-interacting case at $n = 1$ is twice the approximate density of charge carriers in the strongly-interacting case at $n = 1.5 ( p =0.5)$. 
This is also confirmed by Fig. \ref{fig:Noninteracting compressibility}.

\section{Methods for computing the Fermi surface}

All methods used for computing the Fermi surface derive from the Green's function, $G_{\vec{k}}(\tau) = \langle c^\dag_{\vec{k}}(\tau) c_{\vec{k}}(0) \rangle$, calculated for each discrete $\tau \in [0,80]$ imaginary time step.
The section {\it Restructuring of the Fermi surface} describes how different quantities are used to construct the Fermi surface; here we will present how these quantities are calculated using the Green's function.

\subsection{Luttinger's Theorem applied to a free system}

The size for the Fermi surface is proportional to the total particle number $N$, from Luttinger's theorem.
We calculate the non-interacting/tight binding contour in k-space enclosing $N$ particles

\begin{align}
    N= \sum_k{G_{\vec{k}}(0)}.
\end{align}

\subsection{Momentum distribution function (MDF)}

The MDF is calculated from the Green function:

\begin{align}
    n_{\vec{k}} = G_{\vec{k}}(0)
\end{align}
and find the contour, $n_{\vec{k}} = 1/2$, which defines the surface that encloses the filled k-states.

\subsection{Spectral weight at $A_{\vec{k}}(\omega = 0)$}

In the non-interacting and thermodynamic limits the only k-states that have spectral weight at $\omega = 0$ are the states on the Fermi surface.
For the interacting system, we use this property to find the size of the Fermi surface by examining the k-states that have spectral weight at $\omega = 0.$

The relation between the Green's function and the spectral function is described in equation \ref{eq:Spectral function}.
\begin{equation} \label{eq:Spectral function}
    G_{ \vec{k}}(\tau) = \int_{-\infty}^{\infty}{ \frac{e^{-\omega \tau}}{1 + e^{-\beta \omega}} A_{ \vec{k}}(\omega) d\omega}
\end{equation}

We implement an iterative maximum entropy method to calculate the spectral function that most accurately reproduces the input Green's function within error bars.

The Fermi surface is constructed by finding the energy value that maximizes the function $A_{E(\vec{k})}(\omega = 0)$, where $E(\vec{k})$ is the tight binding dispersion: $E(\vec{k}) = -2t(\cos k_x + \cos k_y)$. 
We take this energy to be $E_f$ and create a tight binding contour to approximate the Fermi surface. 

The spectral function at the Fermi wave-vector is constructed by averaging the spectral function for each momentum statistically weighted by $A_{\vec{k}}(\omega = 0)$:

\begin{align}\label{eq:contour}
    A_{kf}(\omega) = \sum_{\vec{k}}{A_{\vec{k}}(\omega) A_{\vec{k}}(\omega = 0)} / \sum_{\vec{k}}{ A_{\vec{k}}(\omega = 0)} 
\end{align}

We construct the spectral function at the Fermi wave vector in this manner because of its relevance in the non-interacting/thermodynamic limit which exactly gives the spectral function at $k_f$. 

\subsection{Retarded Green's function}

\begin{align} \label{eq:Retarded Greens}
    G_{\vec{k}}^{ret}(E) = {\cal P}\int_{-\infty}^{\infty}{d\omega \frac{A_{\vec{k}}(\omega)}{\omega -E}}
\end{align}

The retarded Green's function is calculated using the spectral function in equation \ref{eq:Retarded Greens}, $G_{\vec{k}}^{ret} (E)$. 
We construct the Fermi surface by finding the contour over the Brillouin zone where the Green's function changes sign, or when $G_{\vec{k}}^{ret} (E=0) = 0$. 
When the sign change is pole-like, instead of zero-like, a quasiparticle is present on the Fermi surface for that momentum value.

\bibliographystyle{apsrev4-1}
\bibliography{references}